\documentclass[conference]{IEEEtran}
\IEEEoverridecommandlockouts
\usepackage{cite}
\usepackage{amsmath,amssymb,amsfonts}
\usepackage{graphicx}
\usepackage{textcomp}
\usepackage{xcolor}
\usepackage{subfigure}

\usepackage{algpseudocode}
\usepackage{algorithmicx}
\usepackage[ruled,vlined,linesnumbered]{algorithm2e}
\usepackage[english]{babel}
\usepackage{float}
\usepackage{stfloats}
\usepackage{url}
\usepackage{multirow}
\usepackage{comment}
\usepackage{CJK}
\usepackage{mathrsfs}
\usepackage{enumitem}
\usepackage{physics}
\usepackage[letterpaper, left=0.625in, right=0.625in, top=0.75in, bottom=1in]{geometry}

\IEEEoverridecommandlockouts

\def\BibTeX{{\rm B\kern-.05em{\sc i\kern-.025em b}\kern-.08em
    T\kern-.1667em\lower.7ex\hbox{E}\kern-.125emX}}

\makeatletter

\def\@fnsymbol#1{\ensuremath{\ifcase#1\or *\or \dagger\or \ddagger\or
   \mathsection\or \mathparagraph\or \|\or **\or \dagger\dagger
   \or \ddagger\ddagger \else\@ctrerr\fi}}

\begin{document}

\title{{\fontsize{18}{0}\selectfont Unleashing the True Power of Age-of-Information: Service Aggregation in Connected and Autonomous Vehicles \vspace{-0.25 in}\\}
}


\author{
\IEEEauthorblockN{Anik Mallik\IEEEauthorrefmark{1}, Dawei Chen\IEEEauthorrefmark{2}, Kyungtae Han\IEEEauthorrefmark{2}, Jiang Xie\IEEEauthorrefmark{1}, and Zhu~Han\IEEEauthorrefmark{3}\\
\IEEEauthorblockA{\IEEEauthorrefmark{1}\textit{Department of Electrical and Computer Engineering, The University of North Carolina at Charlotte, NC, USA}}
\IEEEauthorblockA{\IEEEauthorrefmark{2}\textit{InfoTech Labs, Toyota Motor North America R\&D, Mountain View, CA, USA}}
\IEEEauthorblockA{\IEEEauthorrefmark{3}\textit{Dept. of Electrical and Computer Engineering and the Dept. of Computer Science, University of Houston, TX, USA}}\vspace{-0.64 in}}
\thanks{\vspace{-0.02 in}This work was supported by funds from Toyota Motor North America and by the US National Science Foundation (NSF) under Grant No. 1910667, 1910891, and 2025284.}
}


\maketitle

\begin{abstract}

Connected and autonomous vehicles (CAVs) rely heavily upon time-sensitive information update services to ensure the safety of people and assets, and satisfactory entertainment applications. Therefore, the freshness of information is a crucial performance metric for CAV services. However, information from roadside sensors and nearby vehicles can get delayed in transmission due to the high mobility of vehicles. Our research shows that a CAV's relative distance and speed play an essential role in determining the Age-of-Information (AoI). With an increase in AoI, incremental service aggregation issues are observed with out-of-sequence information updates, which hampers the performance of low-latency applications in CAVs. In this paper, we propose a novel AoI-based service aggregation method for CAVs, which can process the information updates according to their update cycles. First, the AoI for sensors and vehicles is modeled, and a predictive AoI system is designed. Then, to reduce the overall service aggregation time and computational load, intervals are used for periodic AoI prediction, and information sources are clustered based on the AoI value. Finally, the system aggregates services for CAV applications using the predicted AoI. We evaluate the system performance based on data sequencing success rate (DSSR) and overall system latency. Lastly, we compare the performance of our proposed system with three other state-of-the-art methods. The evaluation and comparison results show that our proposed predictive AoI-based service aggregation system maintains satisfactory latency and DSSR for CAV applications and outperforms other existing methods.
\end{abstract}


\vspace{-0.1in}
\section{Introduction}
\vspace{-0.05 in}
Recent advancements in connected and autonomous vehicles (CAVs) have enabled vehicles and roadside sensors in the vicinity to share and receive information updates on numerous tasks, such as dynamic map updates, probable collision detection, and obstacle recognition \cite{wang2018networking}. The communication between vehicles and sensors is used to run both safety (e.g., collision, asset damages, and public safety) and entertainment applications (e.g., cooperative extended reality). In these applications, broadcasting by roadside sensors and nearby connected vehicles is a common way to convey information to an ego vehicle (i.e., the vehicle in consideration or the target vehicle) \cite{ma2011design}. The protocol design and security enhancement of such broadcast messages have been well-studied for vehicular ad-hoc networks (VANETs) and CAVs \cite{lyu2017ss}. However, new challenges arise when more sensors and vehicles are connected to a single vehicle. For instance, a CAV has to determine whether and when it needs the information update service from another sensor or vehicle. Additionally, with mobility, the Age-of-Information (AoI) or the freshness of information received from different sources will vary as well, which may significantly impact the CAV application performance. 

\par
\begin{figure}[t!]
\centerline{\includegraphics[scale=0.45]{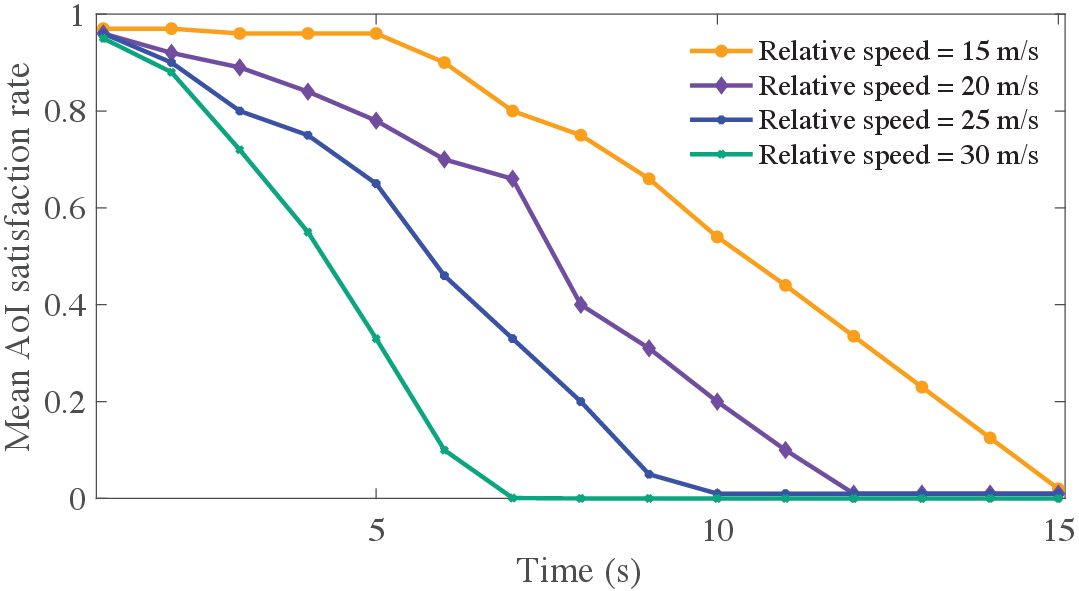}}
\vspace{-0.1 in}
\caption{Mean AoI satisfaction rate of stationary sensors and moving vehicles at different relative speeds of the ego vehicle.}
\label{fig:motiv1}
\vspace{-0.25 in}
\end{figure}

 AoI has been used as a performance metric for low-latency applications, especially in CAVs \cite{zhang2022aoi}. It is defined as the time elapsed between information generation by a source and information reception by a requester. It is important for CAVs to maintain a satisfactory AoI to keep track of the road and nearby objects in real-time to ensure public safety as much as possible. Variations in the mobility of vehicles can increase the AoI of the information received, which may fail to meet the application requirements \cite{xu2022aoi}. Our initial study shown in Fig. \ref{fig:motiv1} indicates that the mean AoI satisfaction rate varies a lot with the relative speed of the ego vehicle with respect to other sensors and vehicles with low coverage areas. AoI satisfaction rate is defined as the percentage of the AoI of the information updates received from stationary sensors and moving vehicles within the upper bound of the required AoI. 

\par In low-latency CAV applications, it is of pivotal importance that the information updates from different sources received by a CAV are sequenced within a short period of time. According to \cite{ghoshal2023performance}, this latency requirement can be as low as $100$ ms. If the received information from different sources is not in the correct sequence, or the expected information update is not received at all from a specific source, then the entire application may not function. For example, a pedestrian's exact location is essential for a CAV to determine its next course of action. Another example can be drawn from entertainment applications where out-of-sequence information received from different sources may bring staleness to extended reality services, which in turn will cause discomfort or nausea to the passengers in a CAV. 

\par The high mobility of vehicles and low coverage areas of roadside sensors make the information update service for CAVs more challenging. Existing technologies (e.g., IEEE 802.11p, WAVE, and DSRC) allow the roadside sensors and vehicles to have coverage areas of $100$m and $300$-$500$m, respectively \cite{federal2018traffic, v2v, kim2014area}. The high mobility of vehicles causes serious data sequencing issues. The broadcast messages from the sensors and vehicles received by the ego vehicle need to be processed sequentially and within the required latency. However, \textit{due to the fast movement of the ego vehicle from one coverage area to another, data may arrive out-of-sequence or not arrive at all. In such high mobility scenarios, the ego CAV needs to quickly determine which information update service should be terminated, maintained, or established.} In this way, data sequencing can be executed more effectively.

\begin{figure*}[t!]
\centering
\includegraphics[width=1\textwidth]{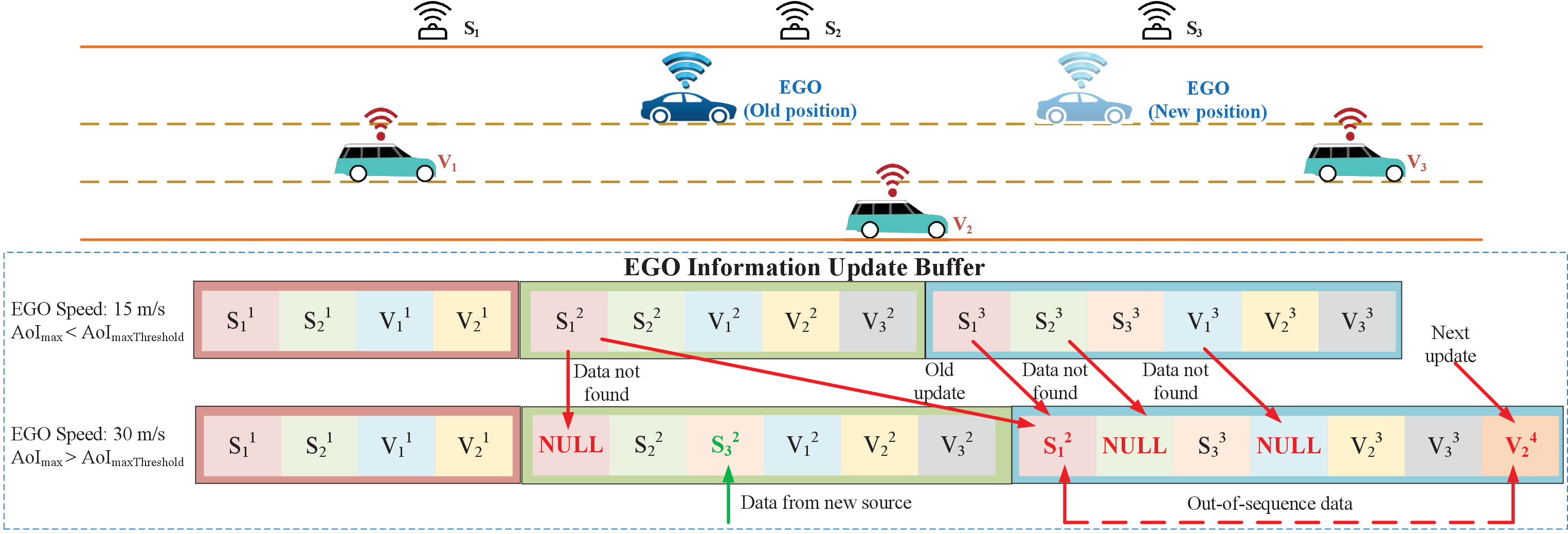}
\vspace{-0.3 in}
\caption{Service aggregation problem for varying mobility of the ego vehicle in a CAV scenario.}
\label{fig:DataSync}
\vspace{-0.25 in}
\end{figure*}

\par We define ``service aggregation'' in CAVs as processing multiple information update services according to their update cycles sequentially from different sensors and nearby vehicles received by an ego vehicle via broadcast messages. We consider that service aggregation involves two tasks: one is service connection (i.e., maintain/terminate/initiate a service), and the other is data sequencing. Note that security is an important aspect of service aggregation in CAVs, which is widely studied in wireless sensor networks (WSNs) \cite{yang2008sdap} and VANETs, and not in the scope of this paper.



\par In this paper, we argue that AoI is so far used as a performance metric for CAVs, while it has the potential to improve the performance of service aggregation in CAVs as well. The question this paper aims to address is, how can we unleash the true power of AoI (i.e., the use of AoI as a tool to enhance the overall system performance)? This paper presents a novel predictive AoI-based service aggregation method for CAVs that can determine highly accurate data processing sequences ahead of time and serve low-latency applications better than existing service aggregation methods in terms of processing delay. To the best of our knowledge, this is one of the first research works on service aggregation in CAV scenarios.

\textbf{Motivations and challenges:} In Fig. \ref{fig:DataSync}, a simulation of CAVs is demonstrated where the ego vehicle is connected to several roadside sensors and nearby vehicles. The ego vehicle is marked as ``EGO'', and other vehicles and sensors are denoted by $V$ and $S$. In this experimental scenario, sensors and vehicles have coverage areas of $100$m and $300$m, respectively. This figure also shows the information update buffer of EGO with three distinct update groups at different update cycles. The data denotation in the buffer denotes the information source number and update cycle number (e.g, $S_2^3$ means the third information update from sensor $S_2$). In the first scenario, EGO is moving toward the coverage area of $S_2$ from $S_1$ and $V_2$ from $V_1$. When the relative speed of EGO is $15$ m/s, and the maximum AoI (considering all sensors and vehicles) is less than the maximum AoI threshold (required latency), the sequence of the received information update is found to be satisfactory, as shown in the figure. In the other scenario, EGO has a new position at a speed of $30$ m/s, where the maximum AoI is higher than the maximum AoI threshold; it is observed that EGO has left the coverage area of $S_1$ completely. Therefore, the second update from $S_1^2$ is missing, although the buffer at EGO is expecting it, causing a delay. At the same time, a new sensor's update for the second cycle, $S_3^2$, arrives at the buffer. At the next update cycle, this becomes even more challenging when EGO leaves the coverage area of $S_2$ and gains a longer relative distance from $V_1$. As a result, the expected update from $S_1^3$, $S_2^3$, and $V_1^3$ are missing in the buffer, old update $S_1^2$ arrives at the buffer, and data from $V_2^4$ arrives earlier from the fourth cycle due to a lower AoI.

\par This simulation study motivates us to study the impact of relative speed and coverage area on AoI and service aggregation for CAVs. However, there are several research challenges. First, the sensors and vehicles are heterogeneous in nature, and so are their coverage sizes, which makes it challenging to model the AoI. Second, since service aggregation in CAVs has not been studied before, new performance metrics need to be defined for this specific research. Third, the low-latency applications in CAVs have unique maximum tolerable processing latency requirements, which makes it very challenging to make a service connection decision within the maximum tolerable processing latency. $N$-step-ahead prediction can be an answer to this problem, but defining $N$ is a challenging task due to the heterogeneity of devices and services. Finally, prediction will consume the computing resources of the ego vehicle and cost even more time than the processing latency. How to reduce the total number of predictions to save computational resources and time is a crucial factor in such system design. Therefore, we advocate that the service aggregation system should adopt a periodic AoI prediction policy that requires less latency and computational resources to meet the overall Quality-of-Service (QoS) requirements of the application.



\textbf{Our contributions:} Our contributions in this paper are summarized as follows:
\vspace{-0.05 in}
\begin{itemize}
    \item \textbf{AoI modeling and prediction:} We model the AoI from heterogeneous sensors and vehicles for the ego vehicle. Then, we compare several prediction models, and based on prediction latency, complexity, and accuracy, we choose the long short-term memory (LSTM) network for predicting the $N$-step-ahead AoI. 
    
    \item \textbf{System design for service aggregation using predictive AoI:} We propose a service connection policy and information update system for highly accurate service aggregation using the predictive AoI. The proposed system determines when to initiate/terminate/maintain a connection and how to aggregate a service in high-mobility CAV scenarios.

    \item \textbf{Performance evaluation and comparison:} We set up a simulation environment to best mimic the real-world complexities using OMNet++. The proposed system's performance is evaluated and compared with three state-of-the-art service aggregation methods. Our proposed system outperforms each of them for varying relative speed of high-mobility CAVs.

\end{itemize}
\vspace{-0.15in}
\section{Related Work}
\vspace{-0.05in}
\textbf{Service aggregation in WSN, VANET, and UAV:} Information fusion or data aggregation is a well-researched topic for WSNs. Numerous information fusion techniques are proposed \cite{nakamura2007information}, such as data aggregation based on clustering and compression \cite{yuea2012energy}. However, these methods do not consider the high mobility of CAVs connected to stationary sensors and moving vehicles, which makes it not viable to implement for CAVs.


\par VANETs and UAVs are also studied in terms of service migration/handoff, data fusion, and task scheduling. Methods for reducing handoff delay or improving the handoff process, and reducing task scheduling load for CAVs and UAVs are proposed \cite{srivatsa2008performance, nasrin2018sharedmec, qi2020scalable}. Nevertheless, task scheduling often refers to prioritizing tasks on the basis of their importance in a connected vehicular network, which does not solve the problem of aggregation of a singular low-latency service for CAVs.

\par \textbf{AoI in connected vehicles:} AoI is recognized as an essential performance metric for CAVs. Researchers propose to minimize AoI through priority-based task scheduling, joint optimization, and machine learning \cite{xu2022aoi, zhang2022aoi}. \textit{However, there is a lack of extensive research on the use of AoI to improve the overall system performance.} In \cite{ni2018vehicular}, AoI is applied to schedule broadcast messages from vehicles to base stations to avoid collisions. Nonetheless, the CAV system considered in our research studies the broadcast service aggregation at a granular level where service connection and aggregation for a single service take place to maintain satisfactory latency using AoI, which has not been researched at all.

\vspace{-0.1in}
\section{System Model} \label{sec:Model}
\vspace{-0.05in}
CAVs are connected to two major types of information sources: roadside sensors and nearby vehicles. Let $S = \{S_1, S_2,...\}$ be the set of sensors and $V = \{V_1, V_2,...\}$ be the set of nearby vehicles. The ego vehicle is denoted as $E_v$ in this paper. $S$ and $V$ broadcast information toward $E_v$ at a specific interval based on the update requirement. The application running on $E_v$ requires $Q$ updates per unit time to meet the refresh rate (i.e., number of refreshes per unit time). Therefore, $E_v$ is supposed to receive information updates at every $1/Q$ interval from $S$ and $V$. If the $E_v$ application runs for a total time, $T$, then the total number of information updates becomes $QT$. The set of information updates over the entire application runtime can be denoted as $U = \{U^1, U^2,..., U^m,..., U^{QT}\}$, where $U^m$ is the update for the $m$th cycle, which is at an arbitrary point in time, $t^m$.

\par Each information update originated from $S$ or $V$ has a unique AoI when received by $E_v$ due to the propagation delay through the wireless medium. Hence, the AoI from a source node $n$ (where $n\in S\cup V$), at update cycle $m$, can be expressed as

\vspace{-0.1in}
\begin{small}
\begin{equation}
t_n^m = T_{mReq}^m + \frac{d_n^m}{c} - T_n^m,
\label{eq:AoIn}
\vspace{-0.08in}
\end{equation}
\end{small}\noindent
where $T_{mReq}^m$ is the time of information requested by $E_v$ for the $m$th cycle's update and $T_n^m$ is the time of information originated by the source node $n$. Here, ${d_n^m}/{c}$ is the propagation delay, where $d_n^m$ is the distance between $E_v$ and the source node $n$ at time $t_m$, and $c$ is the speed of light. Now, the mean AoI for the source node $n$ considering all the information updates, $U$, can be modeled as

\vspace{-0.2in}
\begin{small}
\begin{equation}
A_n = \frac{1}{QT}\sum_{m=1}^{QT}t_{n}^m,
\label{eq:averageAOI}
\vspace{-0.05in}
\end{equation}
\end{small}

\par $E_v$ has an information update buffer where there are update segments for each element of $U$ (refer to Fig. \ref{fig:DataSync}). Each update segment is supposed to contain information updates for the corresponding update cycle only from all the sources that have $E_v$ in their coverage areas. If the update segment for the $m$th cycle receives an update for the $(m\pm p)$th cycle (where $(m\pm p)\in\displaystyle\mathop{\mathcal{Z}}$ and $\mathcal{Z}$ is the set of whole numbers), then this event is considered as a data sequencing issue in this work.

\vspace{-0.05in}
\section{Proposed CAV Service Aggregation using Predictive AoI} \label{sec:Proposed}
\vspace{-0.05in}
In this paper, we assume that the information update messages contain information about the source node (i.e., whether it is a stationary roadside sensor or a mobile vehicle). Moreover, all the connected vehicles exchange basic information with each other, such as speed and geolocation (e.g., latitude-longitude). Thus, the relative speed of the EGO vehicle with respect to other connected vehicles can be derived. 

\vspace{-0.1in}
\subsection{AoI Prediction} \label{sec:PredictAoI}
\vspace{-0.05in}
From our initial study, we observe that there is an implicit relation between the relative speed of the ego vehicle and the AoI from specific information source nodes. The predictive AoI model for each source node, $n$, thus has two input parameters: the timestamp and relative speed of $E_v$ with respect to $n$. The relative speed is a function of the relative distance between $E_v$ and $n$, and time.

\par
\begin{figure}[t!]
\centerline{\includegraphics[scale=0.4]{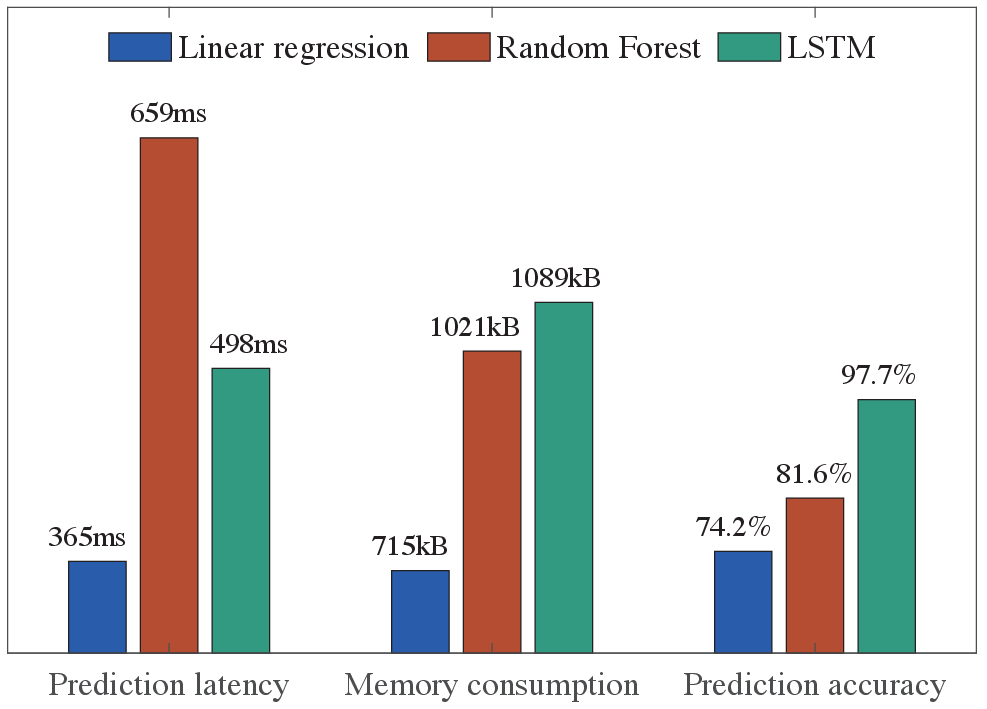}}
\vspace{-0.15 in}
\caption{Comparison of AoI prediction latency, memory consumption, and accuracy of different prediction models.}
\label{fig:PredComp}
\vspace{-0.15 in}
\end{figure}

 For a low latency CAV application, the prediction of AoI should be done in such a way that it does not introduce significant additional load to the system. For instance, if the system has a certain latency requirement for each update, the prediction needs to be completed within a time frame, leaving sufficient time for service aggregation tasks to be done by the required latency. Moreover, the accuracy of prediction is of utmost importance in the case of CAVs due to the involvement of safety issues. Finally, the prediction needs to be done for each source node. With an increase in the number of sources, the computational load for prediction increases -- which emphasizes the use of a low-complexity prediction model. Therefore, before choosing a prediction model, the latency, accuracy, and computing load need to be evaluated first. We implement three prediction models to predict the AoI in CAVs, which are linear regression, Random Forest, and long short-term memory (LSTM) network. Fig. \ref{fig:PredComp} shows the comparison results of the three prediction models. The necessity of a trade-off is evident here since each model has different pros and cons. Being the highest priority performance metric, accuracy and latency dictate the use of the LSTM network in our research. The high accuracy provided by the LSTM network is due to its better and recurrent understanding of the temporal dependencies of our training dataset.

\par
\begin{figure}[t!]
\centerline{\includegraphics[scale=0.42]{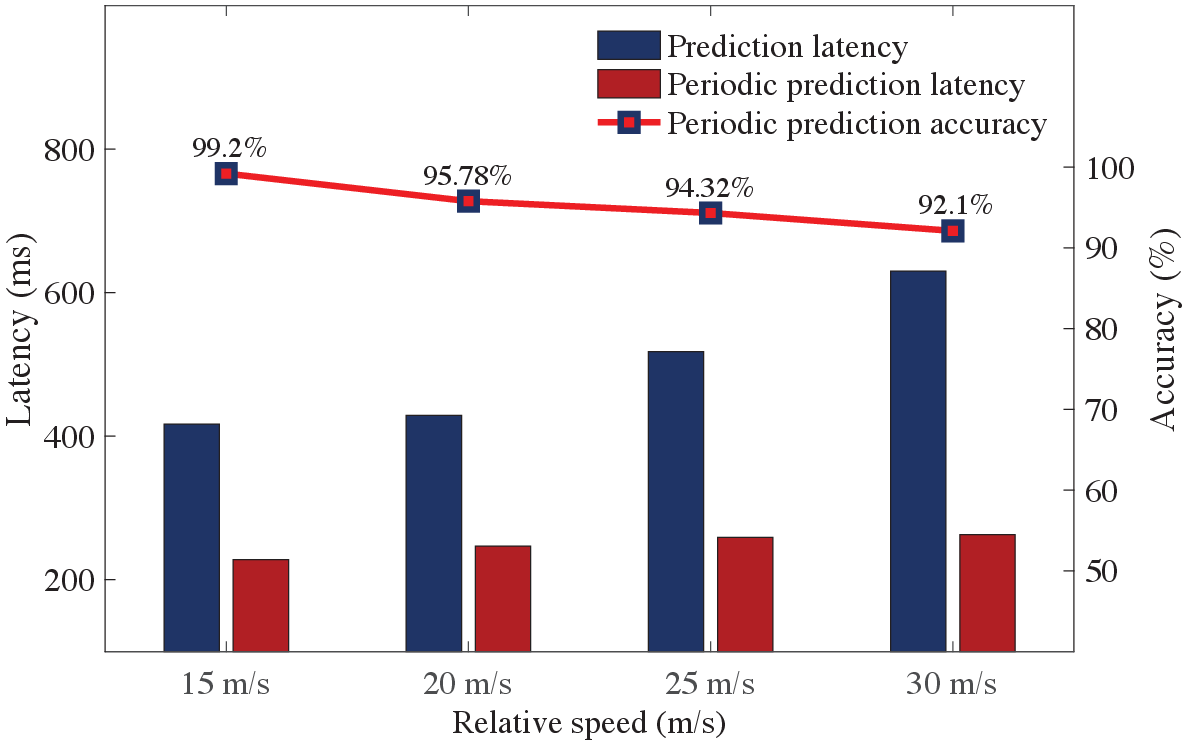}}
\vspace{-0.15 in}
\caption{AoI prediction latency, periodic AoI prediction latency, and periodic AoI prediction accuracy using LSTM network.}
\label{fig:PeriodicPred}
\end{figure}

 Though the LSTM network provides the best prediction performance in our research, it still has a high latency and memory consumption, which may not satisfy the overall QoS requirement of the CAV application (e.g., latency and computing load). Consequently, we propose a periodic prediction system that predicts the AoI at a certain interval. Additionally, this prediction system clusters the AoI data of several sources based on similarity over a specific time period. This reduces the computational load and latency by a high margin since the prediction runs alongside the service aggregation tasks. Let the period of prediction be denoted as $N$. The predictive model predicts AoI at every $N$-step, and it predicts $N$-step-ahead AoI. $N$ can be determined by the application developers in numerous ways that allow them to execute the prediction within the upper bound of the latency requirement. We define a new parameter called ``speed-to-coverage area ratio (SCAR)'', which is the ratio of the relative speed of $E_v$ to the coverage area of a source node, $n$ (source node can be either stationary or moving). The prediction should be done at least once before $E_v$ leaves the coverage area of a source. If the prediction latency is denoted as $L_{pred}$, then the range of $N$ that we recommend can be expressed as
\vspace{-0.08in}
\begin{small}
\begin{equation}
N = \left[\frac{L_{pred}}{1/Q} , \frac{Q}{SCAR_{n}}\right],
\label{eq:Period}
\vspace{-0.07in}
\end{equation}
\end{small}\noindent
where $1/Q$ is the maximum AoI threshold. Fig. \ref{fig:PeriodicPred} shows the performance of unit-step and a $3$-step AoI prediction by an LSTM network. Using a multi-step periodic prediction, our system is able to reduce the overall latency by at least $42\%$. The hyperparameter values and types for the training of the LSTM network are shown in Table \ref{table:hyper}.

\begin{table}[t!]
\vspace{-0.2 in}
\setlength{\tabcolsep}{8pt}
\renewcommand{\arraystretch}{0.9}
\begin{center}
\caption{Training Hyperparameters of LSTM Network for AoI Prediction}
\vspace{-0.1 in}
\begin{tabular}{|l|c|}
\hline
\textbf{Hyperparameters} & \textbf{Values/Type}      \\ \hline \hline
No. of LSTM units in each layer     & $64$                   \\ \hline
No. of LSTM layers                  & $4$                    \\ \hline
Dropout rate                        & $0.1$                  \\ \hline
Recurrent dropout rate              & $0.1$                  \\ \hline
Activation function                 & $tanh$                \\ \hline
Weight initializer                  & glorot\_uniform      \\ \hline
Recurrent weight initializer        & orthogonal           \\ \hline
Training batch size                 & $32$                   \\ \hline
Training epochs                     & $50$                   \\ \hline
Optimizer                           & Adam                 \\ \hline
Learning rate                       & $0.001$                \\ \hline
Loss function                       & mean\_sqaured\_error \\ \hline
\end{tabular}\label{table:hyper}
\end{center}
\vspace{-0.35 in}
\end{table}

\vspace{-0.15 in}
\subsection{Service Aggregation using Predictive AoI} \label{sec:ServiceAggPredictAoI}
\vspace{-0.05 in}
The proposed service aggregation using periodic predictive AoI can be divided into five tasks. First, the system initializes two buffers: the outer one is for receiving updates from information source nodes via broadcast messaging (data buffer), and the inner one is to put updates according to their update cycle (update buffer). Second, if the system receives an update from a new node that does not belong to the node list, $S\cup V$, based on its AoI, the system determines whether to establish a new connection and put this update into the respective segment of the update buffer. Third, the system checks whether it is the period for prediction, and if it is, then the LSTM network predicts $N$-step-ahead AoI using the current relative speed, $v$, and timestamp, $t$. If the $N$-step-ahead AoI is satisfactory, then it maintains the service connection at that cycle; otherwise, it terminates the connection with the specific node. The information update is also processed accordingly. Fourth, the system clusters nodes based on the predictive AoI to reduce the computing load. Finally, the system updates the node list, $S\cup V$, and the information update set, $U$. This entire process is illustrated in Algorithm \ref{alg:proposed}.


\section{Performance Results and Discussion}\label{sec:results}
\vspace{-0.05in}
In this section, the experimental setup, performance metrics, performance evaluation of our proposed CAV service aggregation using predictive AoI, and comparison of the proposed system with state-of-the-art methods are discussed.

\begin{algorithm}[bt!]
\DontPrintSemicolon
\footnotesize
\SetNoFillComment
\caption{Proposed CAV service aggregation using periodic predictive AoI.}
\label{alg:proposed}
\KwIn{Relative speed $v_n^m$, Timestamp $t^m$, and $AoI_n^m$.}
\textbf{Begin}\\
\textbf{Initialize} Hyperparameter settings, data buffer, $B$, update buffer, $U$, and period of prediction, $N_n$.\\
\ForEach{update cycle, $m$}{
  \If{$n$ is not in node list}{
    \If{$AoI_n^m\leq AoI_{\text{max}}$}{
      Add $n$ to node list, $S\cup V$\\
      Add $U_n^m$ to $U$\\
    }
    \Else{
      Discard $U_n^m$\\
    }
  }
  \Else{
    \ForEach{source node, $n$}{
      \If{$m \mod N_n = 0$}{
        Input of LSTM $\gets v_n^m$, $t^m$\\
        Output of LSTM: $AoI_n^{m+N_n}$\\
        \If{$AoI_n^{m+N_n} > AoI_{\text{max}}$}{
          Terminate connection with $n$ at update cycle, $m+N_n$\\
          Update node list, $S\cup V$ \\
        }
        \Else{
          Maintain connection with $n$ at update cycle, $m+N_n$\\
          Determine update segment, $U$, for $data_n$\\
          Put $U_n$ to selected update segment, $U$\\
        }
      }
      \Else{
        \If{$AoI_n^m \leq AoI_{\text{max}}$}{
          Add $U_n^m$ to $U$\\
          Maintain connection with $n$ at update cycle, $m$\\
          Cluster $n$ with other nodes with equal $AoI^m$\\
          Update node list, $S\cup V$ with cluster nodes\\
        }
        \Else{
          Discard $U_n^m$\\
        }
      }
    }
  }
}
\KwRet {node list, $S\cup V$, and information update set, $U$}\\
\KwOut{Service connection decision and information update, $U_n^m$.}
\end{algorithm}

\vspace{-0.1 in}
\subsection{Experimental Setup} \label{sec:detailsExperiment}
\vspace{-0.05in}
We conduct extensive experiments using simulated CAV scenarios with OMNet++. For the mobility model of the CAVs, a modified version of the freeway model is used from \cite{arbabi2010highway}. The speed of the CAVs ranges from $15-30$ m/s. The coverage areas of sensors and vehicles are set to be $100$m and $300$m, respectively. Each simulation lasts for $20$ minutes. The prediction periods for sensors and vehicles are set to be $5$ and $10$. The training and testing dataset used for the LSTM network contains data of $96,000$ and $12,000$ timestamps, respectively. We set the required information update frequency to $3$ updates per second, which makes the maximum AoI threshold $333$ ms.

\vspace{-0.1in}
\subsection{Performance Metric}
\vspace{-0.05in}
We evaluate and compare the proposed system's performance in terms of the overall latency and data sequencing success rate (DSSR) at different relative speed values of the ego vehicle and continuously varying relative speeds at different timestamps. DSSR is introduced in this research for the first time as a performance metric of CAV service aggregation, which is defined as the percentage of successful data sequencing with respect to the total number of data in a buffer segment.

\vspace{-0.05in}
\subsection{Overall Latency of the Proposed System}
\begin{figure}[t!]
\subfigure[]
{\includegraphics[width=0.28\textwidth]{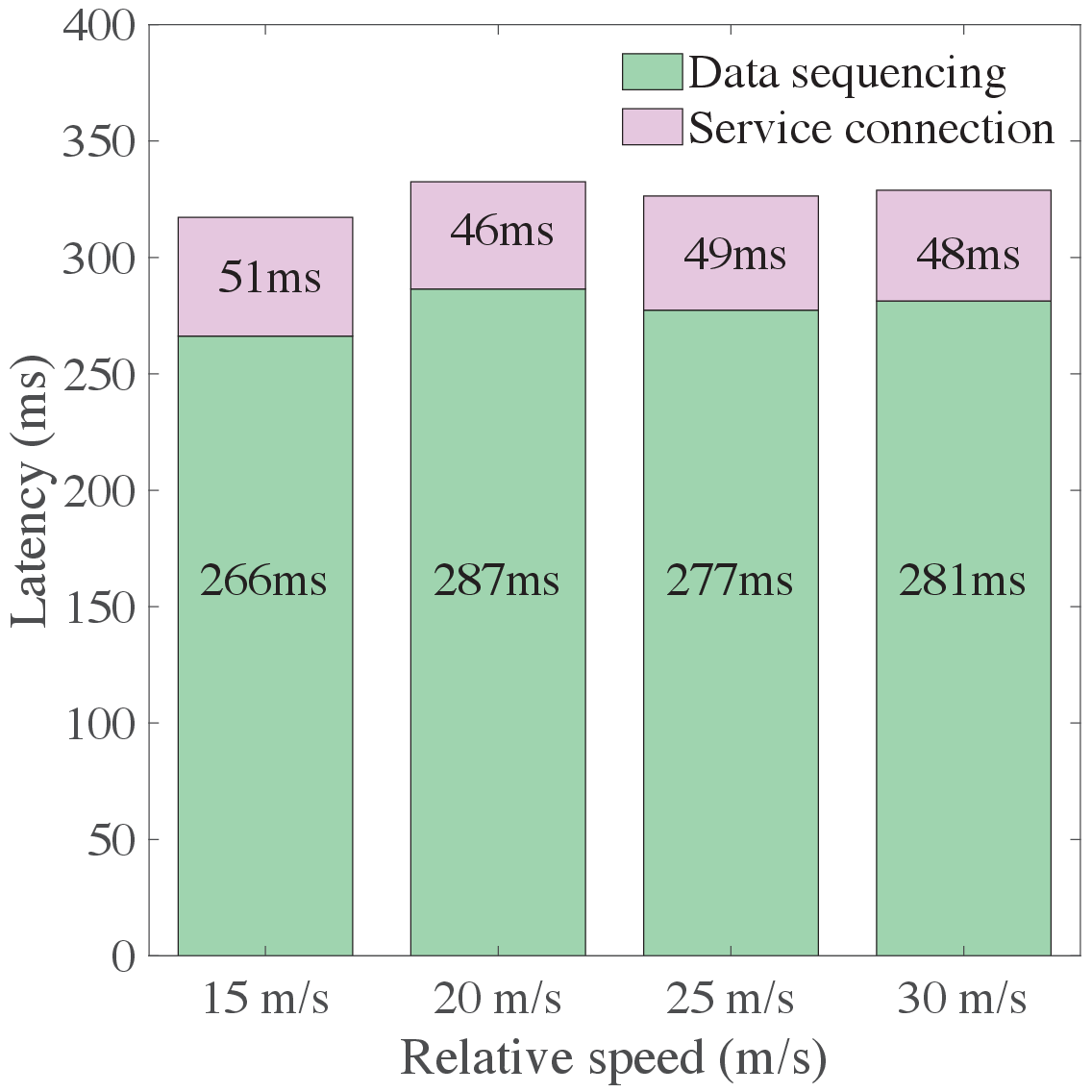}\label{fig:latSystemBar}}
\hspace*{\fill}
\subfigure[]
{\includegraphics[width=0.19\textwidth]{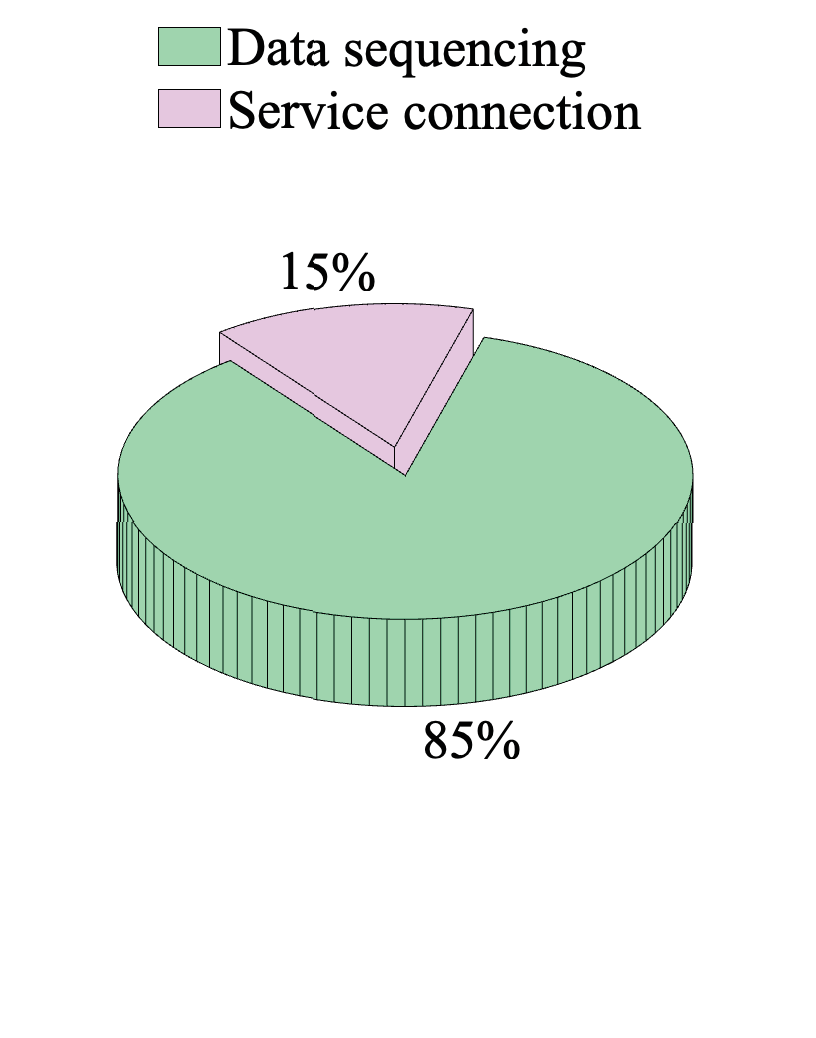}\label{fig:latSystemPie}}
\vspace{-0.15 in}
\caption{(a) Mean system latency at different relative speed and (b) overall latency of system tasks.}
\label{fig:latSystem}   
\vspace{-0.15 in}
\end{figure}

\vspace{-0.05in}
The overall system latency of the proposed CAV service aggregation can be divided into two parts: data sequencing and service connection. The mean overall latency at different relative speeds of the ego vehicle and the components of the system are shown in Fig. \ref{fig:latSystem}. The mean overall latency of the system is around $326$ ms, whereas data sequencing takes around $85\%$ of the total latency.

\subsection{Proposed System's Performance at Varying Speeds}

\begin{figure}[hbt!]
\centerline{\includegraphics[scale=0.45]{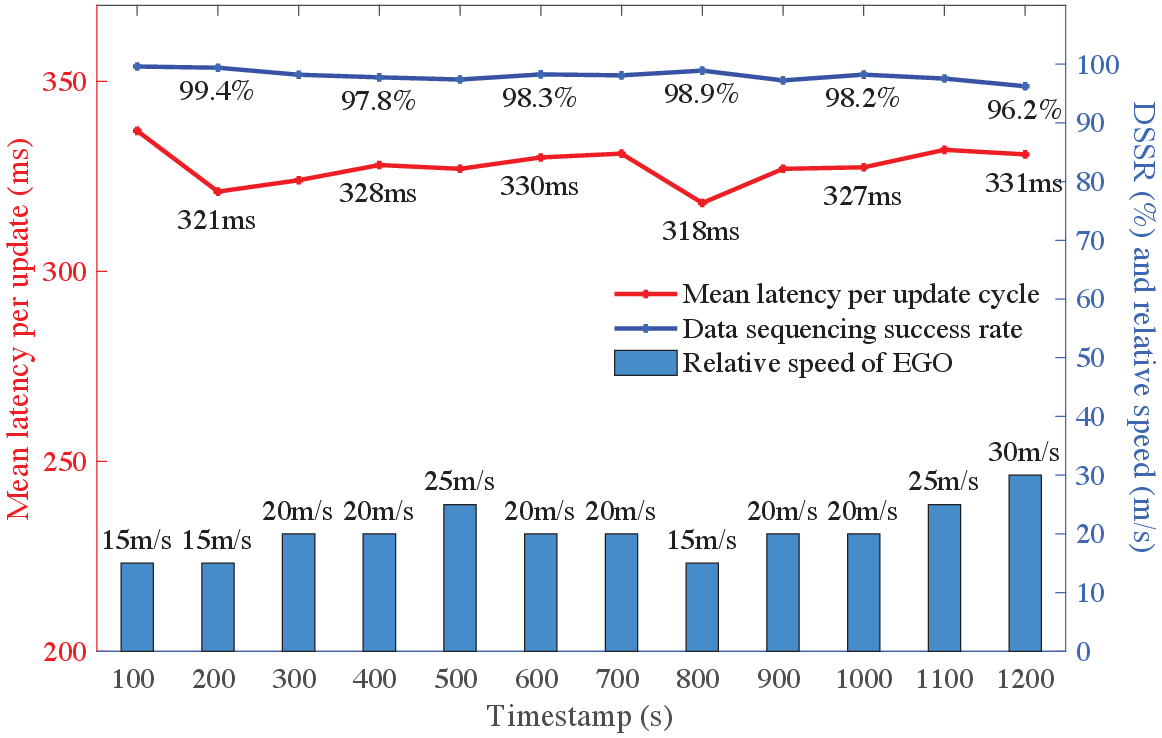}}
\vspace{-0.15 in}
\caption{Mean system latency and data sequencing success rate (DSSR) at varying relative speed and timestamps.}
\label{fig:PredAoICont}
\vspace{-0.1 in}
\end{figure}

In Fig. \ref{fig:PredAoICont}, the proposed system's performance is shown at continuously changing relative speeds of the ego vehicle in terms of mean latency per update and DSSR. It is observed that the system is able to maintain a mean overall latency of $327$ ms with a DSSR of $98\%$ considering all the relative speeds. The initial latency (at $100$ ms timestamp) of the system is a bit higher (around $337$ ms) since the AoI prediction is executed at a later timestamp. Note that the highest priority of our proposed system is to maintain a latency that is under the maximum tolerable threshold ($333$ ms in this case), which may come at a cost of slightly reduced DSSR at a higher relative speed.

\vspace{-0.15in}

\vspace{-0.05in}
\subsection{Comparison of System DSSR and Latency}
The performance of the proposed CAV service aggregation using periodic predictive AoI (denoted as ``Proposed System'' in this subsection) is compared with three other state-of-the-art data communication and queuing techniques listed below. Since there are no existing service aggregation methods available, we modify these methods to fit the research problem of service aggregation.
\begin{itemize}
    \item \textbf{FIFO:} First-in-first-out (FIFO) is a common data queuing and dequeuing method, where the data are served first that arrive first.
    \item \textbf{Stop-N-Wait:} This is a popular data link and transport layer protocol for data communication, which is modified for this research to stop and wait for the data in the correct information update cycle.
    \item \textbf{Priority Queue:} This is another queuing technique that assigns different priorities to data and serves according to the pre-set priorities. In this paper, we set a higher priority to the information updates from nearby vehicles.
\end{itemize}

\par
\begin{figure}[hbt!]
\centerline{\includegraphics[scale=0.42]{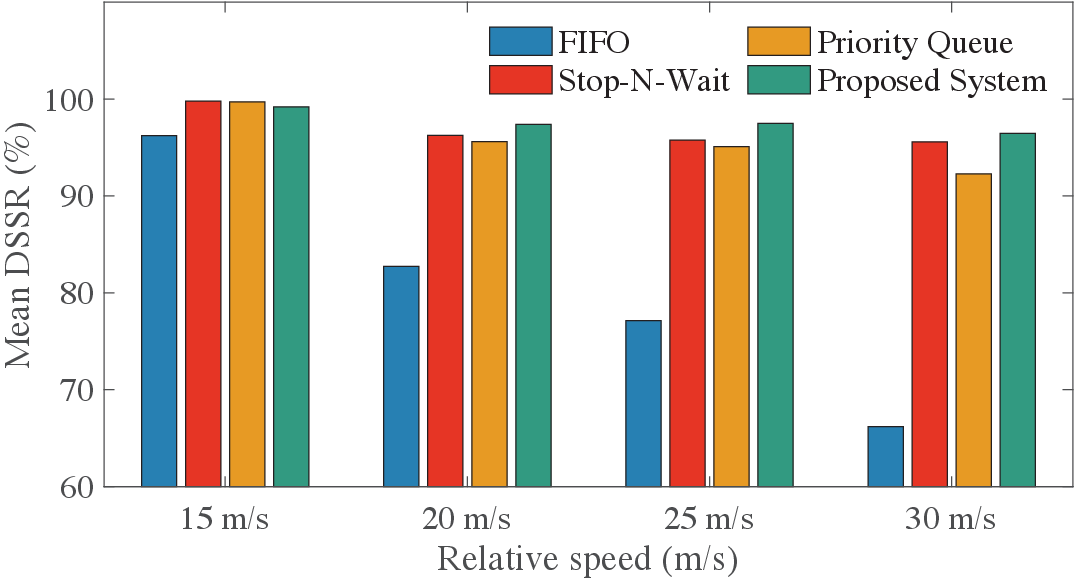}}
\vspace{-0.15 in}
\caption{Comparison of mean data sequencing success rate per update cycle.}
\label{fig:CompDSSR}
\vspace{-0.1 in}
\end{figure}

\begin{figure}[htb!]
\centerline{\includegraphics[scale=0.42]{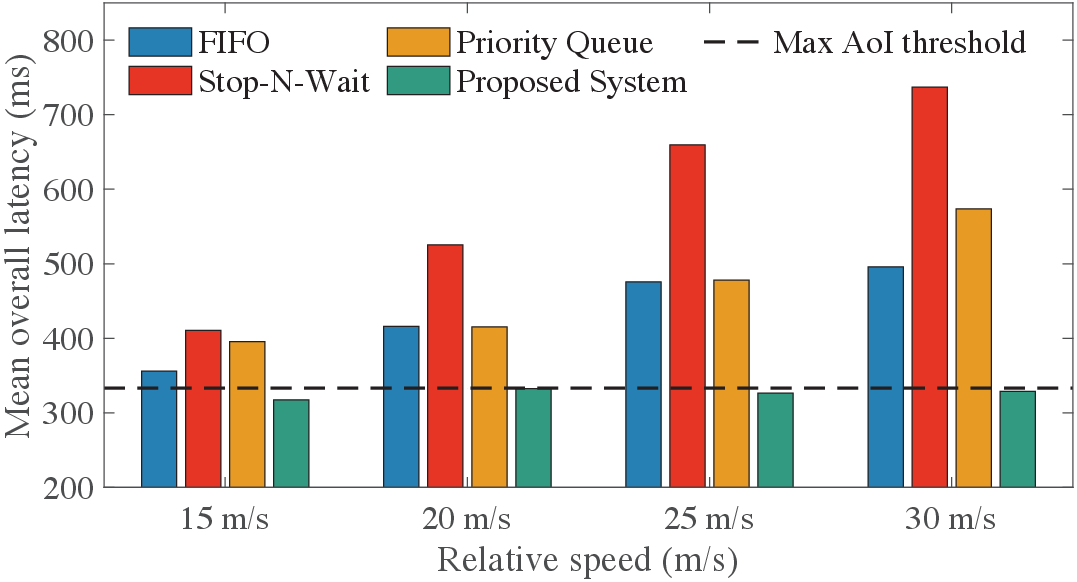}}
\vspace{-0.15 in}
\caption{Comparison of mean overall latency per update cycle.}
\label{fig:CompLat}
\vspace{-0.15in}
\end{figure}

 The comparison among the four service aggregation methods is shown in terms of mean DSSR (\%) and mean overall latency (ms) per update at different relative speeds of the ego vehicle in Fig. \ref{fig:CompDSSR} and \ref{fig:CompLat}, respectively. DSSR in FIFO declines sharply with an increase in speed since the ego vehicle passes the coverage area but does not terminate the connection or adjust its data buffer accordingly. Priority queue also experiences a declining DSSR with an increase in speed due to a similar effect. An interesting finding here is that the DSSR is almost same in case of Stop-N-Wait and Predictive AoI in every relative speed because both methods wait for the correct information update. However, the mean latency is much higher in Stop-N-Wait because of the high waiting time. Predictive AoI can achieve $78\%$ lower latency due to its periodic prediction and clustering.
\vspace{-0.15in}

\vspace{-0.15in}

\par \textit{The average increase in DSSR is $7\%$ and decrease in latency is $51\%$ for the proposed predictive AoI-based CAV service aggregation system compared to the other three methods. Additionally, the predictive AoI-based system successfully maintains the average AoI below the maximum AoI threshold.}

\vspace{-0.05in}
\section{Conclusion}
\vspace{-0.05in}
In this paper, we proposed a novel service aggregation system for time-sensitive CAV applications based on predictive AoI. Our initial study indicated that due to the low coverage areas of connected roadside sensors and nearby vehicles to a CAV, the high mobility of the CAV causes severe degradation in AoI, which in turn makes service aggregation even more challenging. To address this challenge, we proposed a service aggregation system using AoI prediction at a specific interval and clustering information source nodes based on the predicted AoI. The periodic prediction and clustering of source nodes help reduce the computational load and latency. Simulation results showed that the proposed system is capable of predicting the AoI with a high accuracy and providing high DSSR while maintaining the AoI threshold for low-latency CAV applications. Lastly, the performance comparison of the proposed system with other data sequencing methods showed the superiority of the proposed service aggregation system in high-mobility CAV scenarios. 

\vspace{-0.05in}
\linespread{0.9}

\end{document}